\documentclass[a4paper,11pt]{article}

\begin{document}

\section*{BioSig - An application of Octave}

\author{Alois Schl\"ogl}
\date{Mar 1st, 2006}
Author: Alois Schl\"ogl\newline
Affiliation: Institute for Human-Computer Interfaces, University of Technology Graz, AUSTRIA\newline
email: alois.schloegl@tugraz.at\newline
Subj-class: cs/MS\newline

\subsection*{Abstract}
BioSig is an open source software library for biomedical signal processing. Most users in the field are using Matlab; however, significant effort was undertaken to provide compatibility to Octave, too. This effort has been widely successful, only some non-critical components relying on a graphical user interface are missing. Now, installing BioSig on Octave is as easy as on Matlab. Moreover, a benchmark test based on BioSig has been developed and the benchmark results of several platforms are presented.

\subsection*{Introduction (The past ...) }

BioSig is an open source software library for biomedical signal processing. BioSig was developed as research tools for analyzing the human electroencephalogram (EEG) and other biomedical signal. The main issues in this field are the development of new analysis methods for dedicated purposes like an EEG-based brain computer interface (BCI), or automated sleep analysis, or for investigating the neurophysiologic mechanism within the brain. Other biomedical signals, like investigating the electrocardiogram (ECG) as well as respiration, muscle activity, eye movements, blood pressure, oxygen saturation within the blood etc. are of interest. 

Unlike the classical field of signal processing, the area of biomedical signal processing has to deal with artifacts and noise sources, low signal-to-noise ratios, many different data formats, classification and statistical evaluation, and an almost infinite number of data processing methods. 

In the 1990-ties, the use of Matlab became popular in the field of biosignal processing. However, algorithms for biomedical signal processing were rarely available and almost every new PhD-student had to re-implement the methods of his interest. This was also the experience of the author of this work. The research field of biomedical signal processing was characterized by research groups who proposed there own methods, but the algorithms were only available within each group. The field of commercial products was characterized by commercial equipment providers, who usually tend to close the system as much as possible. Often it was difficult to load the recorded data into an idependent analysis. Only very few vendors disclosed their data format or supported exporting the data in an ASCII format. Furthermore, almost every vendor uses its own data format, hindering data exchange. Of course, these facts did did not really foster the development and validation of new analysis methods. 

In this situation, and supported by the success of open source software in the field of operating systems and server software, it became self-evident that an open source library for biomedical signal processing would be useful for the whole field. Accordingly, the BioSig project was born. Based on the widespread use of Matlab, it was obvious to build BioSig on top of it. However, in order to provide a really free and open library, special effort was undertaken for providing compatibility to Octave \cite{Octave}. All functions are tested for there compatibility with Octave and Matlab, too.

\subsection*{Internals of BioSig}
Nowadays, BioSig consists of several subprojects, it supports different programming languages including Octave/Matlab,C/C++; support for Java and Python are in in progress. In the following, we'll describe \emph{BioSig for Octave and Matlab} (\emph{BioSig4OctMat}). 

\emph{BioSig4OctMat} is subdivided into several modules (see \ref{tab:modules}). Within the module "data formats and storage", a common interface for accessing the various formats including an automated format detection are implemented. It supports reading of 40 and writing of 10 different data formats and can be also used to access various audio formats (e.g. WAV, AIF, SND). The preprocessing module provides tools for triggering the data, as well as for artifact detection, artifact reduction and quality control. The signal processing module includes several specialized biosignal processing functions, but also interfaces to standard signal processing functions and wrapper functions for more complex analyses procedures. The classification module includes classification methods (like linear and quadratic discriminant analysis, and interfaces to several libraries on Support Vector Machines (SVM) in combination with cross-validation procedures to prevent overfitting. The module on evaluation criteria contains several functions for performance metrics as used for in the field of BCI research \cite{Schloegl2006b}. The visualization module contain a simple viewer for biomedical data, as well as a wrapper function to visualize the results of several standard analysis procedures. The interactive viewing and scoring software (SViewer) is based on the graphical user interface of Matlab and can not be used with Octave.  

Moreover, BioSig4OctMat depends on the \emph{Time Series Analysis (TSA) toolbox} \cite{TSA, NaN} and the \emph{NaN-toolbox}, which are also part of the Octave-forge repository \cite{OctaveForge}. The TSA and NaN toolboxes are very useful because these toolboxes are able to data with missing values. Missing values can be caused by artifacts, and are encoded by \emph{not-a-number} (NaN) \cite{IEEE754}.

\begin{table}
\caption{The library of m-functions is organized along the following subtasks or topics.}
\label{tab:modules}
\begin{tabular}{ll}
\hline
(i)    & Data Acquisition                                  \\
(ii)   & Data formats and storage                          \\
(iii)  & Preprocessing                                     \\
(iv)   & Signal Processing and Feature Extraction          \\
(v)    & Classification and Statistics                     \\
(vi)   & Evaluation Criteria                               \\
(vii)  & Visualization                                     \\
(viii) & Interactive Viewer and Scoring                    \\
\hline
\end{tabular}
\end{table}

\emph{BioSig4OctMat} contains also several demonstration examples; as well as a benchmark function for comparing the performance of different platforms. The benchmark functions performs some typical processing steps for calculating the classifier of BCI experiment. First some data is loaded; then several features are extracted; the features are used to compute a classifier; a cross-validation procedure (in this case a leave-one-out-method) is used for validating the classifiers. The benchmark can be used to compare different hardware platforms as well as different versions of Octave and Matlab. The results for several different platforms are shown in Table \ref{tab:benchmark}.

\subsection*{Compatibility between Octave and Matlab}

In order to provide full compatibility, some non-standard Octave functions were needed. If available, the functions from the Octave-Forge repository \cite{OctaveForge} are used (BITAND, BITSHIFT, DATENUM, DATESTR, DATEVEC, ISCHAR, NUMEL, NOW, RAT, REGEXP, SPARSE, STRMATCH, STRCMPI, STRNCMP, STRNCMPI, STRTOK, STRVCAT), 
in some cases the functionality was extended (STRMATCH, STRVCAT) or newly implemented (STR2DOUBLE, STRFIND, ISDIR). In the meantime, several of these functions have been incorporated in standard Octave. Thus, these functions are distributed with \emph{BioSig4OctMat} only for the sake of compatibility to some older Octave versions. 

Moreover, the function STR2NUM was replaced because of security concerns. Within BioSig, the function STR2NUM was heavily used to decode the header information of many file formats. Hence, the data input was from an outside source, which must be considered insecure \cite{Wheeler}. The use of EVAL within STR2NUM could enable running any arbitrary octave command including systems calls using UNIX; for example the input string could contain the value "unix('cp a b')" or "unix('rm *')". If the header information of some file contains such a string, it could be passed on to STR2NUM and would be executed within EVAL. In order to avoid this security problem, STR2NUM was completely replaced by STR2DOUBLE which did not use the EVAL-command but parsed the code. 

Recent versions of Octave support the on-the-fly (de-)compression of file access using the option "z" in FOPEN. It was possible to use this feature within \emph{BioSig4OctMat}. Accordingly, it is possible to open GZip-compressed biomedical signal data in Octave. Considerations to define a compression scheme for biosignal data became less important. 

Recently, some XML-based data formats have been used for saving biomedical signal data. The open source toolbox \emph{XMLTREE} \cite{XMLTREE} is partly useful for reading XML-data into Matlab. Unfortunately, it is incompatible with Octave and the functionality of XMLTREE is incomplete. Therefore, a reliable (validated) XML parser is desirable. 

The attempt to make \emph{BioSig4OctMat} fully compatible to Octave as well as Matlab was widely successful. Only the interactive scoring software (a desirable but not mission-critical component) can not be used with Octave. BioSig demonstrates that also a large-scale project can be programmed in such a way that it can run on Matlab as well as Octave without any code modifications. 

Despite all efforts (installing BioSig on Octave is no more difficult than installing it on Matlab), the user feedback suggests that most BioSig users use it in combination with Matlab; the Octave users are below the perceptibility level.

\subsection*{Performance}

\begin{table}
\caption{Time needed to complete the BioSig benchmark test on different platforms.}
\label{tab:benchmark}
\begin{tabular}{llr}
\hline
Hardware (CPU, clock, cache)          & Software           & Time [s]  \\
\hline
\hline
Pentium III, 867 MHz, 256kB           & Matlab 6.5         &  989 \\
\hline
Athlon XP 1800+, 1533 MHz, 256 kB & Octave 2.9.4+(cvs) & 4080 \\
\hline
Athlon XP 1600+, 1403 MHz, 256 kB & Matlab 6.5         &  595 \\
(same computer)                       & Octave 2.9.4+(cvs) & 2812 \\
\hline
Athlon XP 1900+, 1600 MHz, 256 kB & Matlab 7.1         &  340 \\
\hline
Power5 (rev2.1), 1656 MHz, ?          & Octave 2.9.4+(cvs) & 2516 \\
\hline
Opteron 264, 1994 MHz, 1024 MB    & Octave 2.9.4+(cvs) & 1070 \\
\hline
\end{tabular}
\end{table}

Table \ref{tab:benchmark} shows the time needed to complete the BioSig-benchmark test. Although some machines had more than one CPU, the benchmark test used only one CPU. Therefore, the result represents the single CPU performance. The presented results are not exhaustive, because the number of platforms is rather limited and the costs per CPU and energy consumption have not been considered. Despite, the table shows already some interesting results. First, the highest performance of Octave was obtained with an AMD Opteron CPU. The Power5 CPU was 2.5 times slower, followed by the some older Athlon CPU's. For one platform, both Matlab and Octave was available. In this case the performance between Matlab and Octave differed by a ratio of 1:4.7.

\subsection*{Summary (... and the Future)}

\emph{BioSig4OctMat} is an application that can be used with Octave and is available from Sourceforge. Currently, the monthly download range is in the range of some hundred downloads per month. It's installation on Octave is no more difficult, than using the proprietary counterpart. However, the benchmark tests show inferior results using Octave. Investigating the reasons for this bad benchmark result of Octave was beyond the scope of this work. However, the inferior benchmark results suggest that performance reasons could be an important factor. The speed reduction by a factor of 4 means that for obtaining the same computational power one must increase the effort on hardware, space, power supply and cooling power by the same factors. 

In future, the Biosig development will stay commited to compatibility to Octave; nevertheless it can not omit the support for the proprietary alternative. Therefore, it is crucial to BioSig that the M-files are compatible to both, Octave as well as Matlab. This is also true for the TSA- and NaN-toolbox \cite{TSA, NaN} which are part of Octave-forge \cite{OctaveForge} and are also used by BioSig \cite{BioSig}.

Octave has two main advantages to its competitor, Octave is open source and it supports on-the-fly (de-)compression. Unfortunately, these are not sufficient for many users. In order to enlarge the number of Octave users, several issues must be addressed. First, the most important issue is increasing the computational performance of Octave. BioSig provides an application-based benchmark of a real-world problem. Second, because the use of EVAL within functions like STR2NUM is a security risk, it is recommended to replace the use of the EVAL-function within STR2NUM and perhaps a few other functions, too. Third, the future development of BioSig will require improved XML-support. Unlike, the current XML-support in OctaveForge, it should be able to parse and validate a whole XML-file and convert it into a corresponding struct.

\bibliographystyle{plain}

\end{document}